\documentclass[twocolumn,aps,floats,letterpaper,floatfix,groupedaddress]{revtex4-1}
\usepackage{amsmath}
\usepackage{graphicx}
\usepackage{amsfonts}
\usepackage{amssymb}
\usepackage{bm}
\usepackage{epsfig,float,afterpage}
\usepackage[colorlinks,linkcolor=blue,citecolor=blue,urlcolor=blue]{hyperref}
\usepackage[usenames,dvipsnames]{color}

\newcommand{\beq}{\begin{equation}}
\newcommand{\eeq}{\end{equation}}
\newcommand{\bes}{\begin{subequations}}
\newcommand{\ees}{\end{subequations}}
\newcommand{\bea}{\begin{eqnarray}}
\newcommand{\eea}{\end{eqnarray}}
\newcommand{\ba}{\begin{array}}
\newcommand{\ea}{\end{array}}
\newcommand{\beqn}{\begin{eqnarray*}}
\newcommand{\eeqn}{\end{eqnarray*}}

\newcommand{\f}[2]{\frac{#1}{#2}}

\newcommand{\om}{\omega}

\newcommand{\la}{\langle}
\newcommand{\ra}{\rangle}

\newcommand{\dg}{\dagger}

\def\nn{\nonumber}

\begin{document}
%\title{Fock versus coherent state input in waveguide quantum electrodynamics: transport coefficients, Kerr and cross-Kerr effect}
\title{Single photons versus coherent state input in waveguide quantum electrodynamics: light scattering, Kerr and cross-Kerr effect}
\author{Athul Vinu and Dibyendu Roy}
\affiliation{Raman Research Institute, Bangalore 560080, India}
%\date{\today}
\begin{abstract}
While the theoretical studies in waveguide quantum electrodynamics predominate with single-photon and two-photon Fock state (photon number states) input, the experiments are primarily carried out using a faint coherent light. We create a theoretical toolbox to compare and contrast linear and nonlinear light scattering by a two-level or a three-level emitter embedded in an open waveguide carrying Fock state or coherent state inputs. We identify rules to compare light transport properties, the Kerr, and cross-Kerr nonlinearities of the medium for the two types of inputs. A generalized description of the Kerr and cross-Kerr effect for different types of inputs is formulated to compare the Kerr and cross-Kerr nonlinearity between two photons in these models. 
\end{abstract}
%\pacs{42.79.-e, 42.65.-k, 42.50.-p} 

\maketitle
    {\it Introduction:} A decade of experimental activities has established waveguide quantum electrodynamics (QED) as an emergent research discipline \cite{RoyRMP2017, Gu2017}. One principal aim of this discipline is to explore strong light-matter interactions between a few propagating photons without any cavity along the direction of propagation. The waveguide QED systems promise to overcome many limitations of cavity QED systems for building quantum networks of light \cite{RoyRMP2017}. Many exciting phenomena \cite{RoyRMP2017, Gu2017, Astafiev10a, Abdumalikov10, Hoi12, HoiPRL2013, vanLoo2013, Koshino2013, Mitsch14, Dmitriev2017} as well as fascinating all-optical devices \cite{Dayan2008, Hwang09, AstafievPRL2010, Hoi11, Oelsner2013, Shomroni2014, Hamann2018} have been demonstrated in these systems. These have improved our fundamental understanding of quantum and nonlinear optics and led to higher sensitivity in quantum metrology and sensing.

  Much of the early theoretical proposals in the waveguide QED systems \cite{RoyRMP2017, ShenFanPRL2007, Chang07, Yudson08, Shi09, RoyPRB2010, Zheng10, Witthaut10, Fan10, Longo10, RoyPRL2011, Zheng13a, RoyPRA2014, Fang15, Xu15} are utilizing single-photon and two-photon Fock state (photon number states) inputs. Coherent state inputs have also been investigated in some theoretical studies \cite{Zheng10, Koshino2012, Peropadre13, KoshinoNJP2013, Caneva15, RoyPRA2017, Manasi2018, Vinu2020}. However, the experimental studies \cite{Hwang09, Abdumalikov10,  Hoi11, HoiPRL2013, vanLoo2013} in these systems predominately apply a weak light beam in a coherent state to explore the physics of single- or few-photon scattering from single or multiple emitters. It is, therefore, in many cases challenging to compare results for different types of inputs. It also remains unclear if there will be any fundamental difference in those studied phenomena if a true (antibunched) single-photon source is applied in comparison to a weak coherent state! In this paper, one of our goals is to develop a theoretical toolbox (identify a set of rules) to compare linear and nonlinear scattering properties of light in the waveguide QED systems for Fock-state and weak/attenuated coherent-state inputs.

We primarily consider the scattering of one or two light beams by a two-level emitter (2LE) or a three-level emitter (3LE) embedded in an open waveguide. The input beams consist of either single or two Fock state photons or a weak coherent state. We mainly investigate the linear and nonlinear (coherent and incoherent) scattering of input beams and nonlinear interactions between photons of a single beam (Kerr effect) and multiple beams (cross-Kerr effect) generated by correlated scattering. The main results are the following: (a) we identify a set of rules to compare the linear and nonlinear transport properties of single photons and faint coherent state input, and (b) we formulate a generalized description of the Kerr and cross-Kerr effect for different types of inputs, and (c) we compare the Kerr nonlinearity between two single photons by a 2LE to the cross-Kerr nonlinearity between them by a 3LE. Below we explain our results in detail.

{\it Light scattering by 2LE:} We first consider a 2LE side-coupled to a one-dimensional continuum of photon modes inside an open waveguide. The difference in energy between the excited level $|2\ra$ and ground level $|1\ra$ of the emitter is $\hbar \om_{21}$. Our calculation for the scattering of weak coherent state inputs is within the Heisenberg picture of quantum mechanics; we then consider photon modes in momentum space and evaluate the time evolution of operators. The scattering of single photons in Fock state is derived within the Schr{\"o}dinger picture, and it is then convenient to take a real-space description of photon modes. Within the Schr{\"o}dinger picture, the operators are time independent, and we find scattering states of the entire system using   scattering theory. The photon modes in real space are related by the Fourier transform to those in the momentum space. The Hamiltonian of the whole system with a linearized energy-momentum dispersion (e.g., $\om_k=v_g k$) of photons reads in momentum space as \footnote{We ignore losses such as pure dephasing or nonradiative decay for simplicity. However, they can easily be incorporated by following methods in Refs.~\cite{RoyRMP2017}.}:
\bea
\f{\mathcal{H}^k_2}{\hbar}&=&\om_{21} \sigma^{\dg}\sigma+ \sum_k \Big(v_gk \big(a_{k}^{\dg}a_{k}-b_{k}^{\dg}b_{k}\big)\nn\\&+&g_p\sigma^{\dg}(a_{k}+b_{k})+g_p(a_{k}^{\dg}+b_{k}^{\dg})\sigma \Big),\label{Ham2k}
\eea
where $\sigma^{\dg}\equiv |2\ra \la 1|~(\sigma \equiv |1\ra \la 2|)$, and $v_g$ is the group velocity of photons. Here, $a_{k}^{\dg}~[b_{k}^{\dg}]$ is the creation operator of right-moving [left-moving] photon modes. Within the rotating-wave approximation and the dipole approximation (a linear light-matter interaction), the coupling strength of the photon modes with the 2LE is given by $g_p$. To obtain a real-space description of propagating photons at position $x \in [-\mathcal{L}/2,\mathcal{L}/2]$, we define $\tilde{a}_{x}(t)= \sum_k\:e^{ikx}a_{k}(t)/\sqrt{\mathcal{L}}$ and $\tilde{b}_{x}(t)=\sum_k\:e^{ikx}b_{k}(t)/\sqrt{\mathcal{L}}$. Here, we take $\mathcal{L}$ being length of the waveguide which can also be considered our quantization length. %Here, the photon operators at $x<0$ and $x>0$ denote respectively the incident and scattered photons, and the photons at $x=0$ are coupled to the emitter.
Thus, we get the following real-space version of the full Hamiltonian:
\bea
\f{\mathcal{H}^x_2}{\hbar}&=&\om_{21} \sigma^{\dg}\sigma-iv_g\int_{-\mathcal{L}/2}^{\mathcal{L}/2}dx \big(\tilde{a}^{\dg}_x\partial_x\tilde{a}_x-\tilde{b}_x^{\dg}\partial_x\tilde{b}_x\big)\nn\\&+&\bar{g}_p\sigma^{\dg}(\tilde{a}_{0}+\tilde{b}_{0})+\bar{g}_p(\tilde{a}_{0}^{\dg}+\tilde{b}_{0}^{\dg})\sigma \Big),\label{Ham2r}
\eea
where $\bar{g}_p=\sqrt{\mathcal{L}}g_p$.

We take incident light in the right-moving channel injected from the left of the emitter. The coherent state input $|E_p,\om_p\ra$ is a monochromatic, continuous-wave beam of frequency $\om_p~(\om_p=v_gk_p)$ and amplitude $E_p$ (which we here assume to be real). It is an eigenstate of $a_k$: $a_k(t_0)|E_p,\om_p\ra=(\sqrt{\mathcal{L}}E_p/v_g)\delta_{k,\om_p/v_g}|E_p,\om_p\ra$, where $t_0$ is an initial time before the interaction of the input beam with the emitter. The intensity (total number of photons per unit length) of the incident coherent beam is $I_{cp}=\la E_p,\om_p| \sum_{k} a_{k}^{\dg}(t_0)a_{k}(t_0)|E_p,\om_p\ra / \mathcal{L}=E_p^2/v_g^2$. The single- and two-photon Fock state input with a wave-vector $k_p$ are respectively:
\bea
|k_p\ra&=&\f{1}{\sqrt{\mathcal{L}}}\int_{-\mathcal{L}/2}^{\mathcal{L}/2}dx\:e^{ik_px}\tilde{a}_{x}^{\dg}|\varphi\ra, \nn\\
|\bm{k}_p\ra&=&\f{1}{\mathcal{L}}\int_{-\mathcal{L}/2}^{\mathcal{L}/2}\int_{-\mathcal{L}/2}^{\mathcal{L}/2}dx_1dx_2\:e^{ik_p(x_1+x_2)}\f{1}{\sqrt{2}}\tilde{a}_{x_1}^{\dg}\tilde{a}_{x_2}^{\dg}|\varphi\ra,\nn
\eea
where $|\varphi\ra$ denotes the vacuum of the electromagnetic fields. The intensity of a single-photon and a two-photon incident beam are respectively $I_{1p}=\la k_p|\sum_k a_{k}^{\dg}a_{k}|k_p\ra/\mathcal{L}=1/\mathcal{L}$ and $I_{2p}=\la \bm{k}_p|\sum_k a_{k}^{\dg}a_{k}|\bm{k}_p\ra/\mathcal{L}=2/\mathcal{L}$. We assume the emitter in the ground state at $t_0$. We evaluate the time evolution of the incident coherent state and the emitter using the Heisenberg equations following Refs.~\cite{Koshino12, RoyPRA2017}. The finding of outgoing scattering states for the single and two-photon inputs is carried out following Refs.~\cite{ShenFanPRL2007,RoyPRB2010, Zheng10}. The details of the  calculation in both cases are given in the Supplemental Materials \cite{sm}.

To quantify linear and nonlinear light scattering, we calculate transport properties such as the reflection and the Kerr and cross-Kerr phase shifts of the transmitted photon(s). For a side-coupled emitter, the reflection of light is a measure for the transfer of photons from the incident right-moving mode(s) to the left-moving mode(s). So, we define reflection current as 
\bea
\mathcal{J}=i\bar{g}_p\la (\sigma^{\dg}\tilde{b}_0- \tilde{b}^{\dg}_0\sigma) \ra, \label{refcur}
\eea
where $\la \dots \ra$ within the Schr{\"o}dinger picture is an expectation in the full states (e.g., $|k_p^+\ra$ and $|\bm{k}_p^+\ra$ in \cite{sm}) of $\mathcal{H}^x_2$ after the scattering of incident light by the 2LE, and the operators in Eq.~\ref{refcur} are time-independent. For the Heisenberg picture used in coherent state input, the expectation is carried out in $|E_p,\om_p\ra$ but the operators in Eq.~\ref{refcur} are evolved to a time which is much later after the scattering by 2LE takes place. We call reflection current for single-photon and two-photon input respectively by $\mathcal{J}_1$ and $\mathcal{J}_2$, which are 
\bea
\mathcal{J}_{1}&=&\f{v_g}{\mathcal{L}}\f{4\Gamma_p^2}{\Delta_p^2+4\Gamma_p^2}=v_gI_{1p}\f{4\Gamma_p^2}{\Delta_p^2+4\Gamma_p^2},\label{ref1}\\
\mathcal{J}_2&=&v_gI_{2p}\f{4\Gamma_p^2}{\Delta_p^2+4\Gamma_p^2}-(v_gI_{1p})^2\f{16\Gamma_p^3}{(\Delta_p^2+4\Gamma_p^2)^2},\label{ref2}
\eea
where the relaxation rate $\Gamma_p=\bar{g}_p^2/(2v_g)$, and the detuning $\Delta_p=\om_p-\om_{21}$. The reflection coefficient of a single photon  is $\mathcal{R}_{1p} \equiv  \mathcal{J}_1/v_gI_{1p}=|r_{1p}|^2=4\Gamma_p^2/(\Delta_p^2+4\Gamma_p^2)$, which is one for a resonant photon, i.e., $\Delta_p=0$. The first part of $\mathcal{J}_2$ gives an independent reflection of two individual photons by the 2LE, and the reflection coefficient of this process is the same as $\mathcal{R}_{1p}$. The second part of $\mathcal{J}_2$ denotes the correlated reflection of two photons by 2LE, and its strength for resonant photons decreases with increasing light-matter coupling $\bar{g}_p$. While the probability of a photon inside the waveguide interacting with the 2LE is an order of $1/\mathcal{L}$, that for two photons simultaneously interacting with the 2LE is an order of $1/\mathcal{L}^2$. Therefore, the correlated scattering of two photons is smaller than the individual photon reflection by one order of $\mathcal{L}$. While the strength of correlated scattering of order $1/\mathcal{L}^2$ grows with the increasing number of incident photons, there also appear higher order terms of $1/\mathcal{L}^m$ with $m>2$ in correlated reflection of $m$ photons for a finite-length waveguide.

The reflection current for a coherent state input is found to be:
\bea
\mathcal{J}_{c}&=&\f{2\Gamma_p \Omega_p^2}{\Delta_p^2+4\Gamma_p^2+2\Omega_p^2}, \label{refc} \\
&=&v_g I_{cp}\f{4\Gamma_p^2}{\Delta_p^2+4\Gamma_p^2}-(v_g I_{cp})^2\f{16\Gamma_p^3}{(\Delta_p^2+4\Gamma_p^2)^2}+\mathcal{O}(\Omega_p^6), \nn
\eea
where $\Omega_p=\bar{g}_pE_p/v_g$ is the Rabi frequency of the incident coherent state. Thus, $I_{cp}=\Omega_p^2/(2v_g\Gamma_p)$. For a faint coherent state, we find $\mathcal{R}_{1p}$ matches to $\mathcal{R}_{cp} \equiv \mathcal{J}_{c}/v_gI_{cp}$ when we drop $2\Omega_p^2$ from the denominator of $\mathcal{J}_{c}$ in Eq.~\ref{refc} in the limit $\Omega_p \to 0$. %\footnote{Relating linear reflection of a faint coherent state with single-photon reflection leads to $I_{1p}=I_{cp}$ or $E_p^2\mathcal{L}/v_g^2=1$ which indicates that the average photon number for coherent state input is one.}. 
We further identify the correlated reflection contribution in $\mathcal{J}_2$ from $\mathcal{J}_{c}$ by expanding it up to the order of $\Omega_p^4$. The above analysis can be generalized to relate the reflection current $\mathcal{J}_m$ for $m$ number of Fock photons with $\mathcal{J}_{c}$ through its expansion up to $\Omega_p^{2m}$. We have extended the Fock state analysis to $m=3$ to confirm the above generalization. Thus, we could point out a set of rules to compare the linear and nonlinear transport properties for a single- or multi-photon Fock state and coherent state input.

{\it Kerr effect:} The nonlinear scattering of an input beam can be characterized by the so-called optical Kerr effect, in which the refractive index of any optical medium depends on the beam's intensity. In such a case, the refractive index $n$ can be separated in linear and nonlinear parts as \cite{Boyd08} $n=n_0+\bar{n}_2E_p^2$, where $n_0$ is the weak-beam (or single-photon) linear part of the refractive index, and $\bar{n}_2$ is a coefficient representing the nonlinear refractive index. The linear and nonlinear refractive indices are proportional to the linear and nonlinear susceptibilities. For light scattering by a single emitter inside the waveguide, we can relate the complex susceptibility of the medium to the change in phase $\phi_p=\phi_p^{(1)}+\phi_p^{(2)}$ of coherently scattered photons where $\phi_p^{(1)}$ is the linear change in phase for a weak beam (or a single photon) and $\phi_p^{(2)}$ is the nonlinear (two-photon) contribution. Thus, we have $\Delta n=n-n_0 \propto \phi_p^{(2)}$. In the regime of recent experimental interest with few photons \cite{Astafiev10a, Hoi12, HoiPRL2013}, we can write an approximate relation to define the Kerr coefficient $K$ as $\phi_p^{(2)} \equiv \phi_p-\phi_p^{(1)} = KE_p^2$.

For a coherent state input, $\phi_p$ can be computed from the coherent transmission amplitude ${\tilde t}_p$: 
\bea
    &&{\tilde{t}}_p=\f{\la E_p,\om_p| \tilde{a}_{x>0}(t)|E_p,\om_p\ra}{\la E_p,\om_p| \tilde{a}_{x>0}(t)|E_p,\om_p\ra_{g_p=0}}=1+2i\boldsymbol{\chi}(t) \nn\\
    &&=1-\f{2i\Gamma_p}{\Omega_p}e^{i\omega_p(t-\f{x}{v_g}-t_0)}\la E_p,\om_p|\sigma(t-\f{x}{v_g})|E_p,\om_p\ra,
\eea
where $\la \dots \ra_{g_p=0}$ denotes no coupling between the 2LE and the input beam. Here, $\boldsymbol{\chi}(t)$ represents the optical susceptibility of the medium, which includes both linear and nonlinear parts of the susceptibility. The phase $\phi_{p}$ associated with ${\tilde t}_p$ is $\phi_{p}(t)=\tan^{-1}(2{\rm Re}\boldsymbol{\chi}(t)/(1-2{\rm Im} \boldsymbol{\chi}(t)))$.
%\bea
%\phi_{p}(t)=\tan^{-1}{\Big(\f{2\:{\rm Re} \boldsymbol{\chi}(t)}{1-2\:{\rm Im} \boldsymbol{\chi}(t)}\Big)}. \label{presponse}
%\eea
We can get $\phi_p^{(1)}$ from $\phi_p$ by taking $\Omega_p \to 0$, and extract $\phi_p^{(2)}$ at any arbitrary $\Omega_p$ using $\phi_{p}^{(2)}=\phi_{p}|_{\Omega_p\ne 0}-\phi_{p}|_{\Omega_p \to 0}$. However, it is not clear how to define such transmission amplitude for a two-photon Fock state input since the scattered states then have amplitudes of two transmitted photons as well as one transmitted and one reflected photons. Instead, we use the first-order correlation function $G^{(1)}(x',x;t)=\la \tilde{a}^{\dg}_{x'}(t)\tilde{a}_{x}(t)\ra$ to obtain the change in phase. For a coherent state input, we find at $x>0$ and $x'<0$: $G^{(1)}_c(x',x;t)=I_{cp} e^{i\om_p(x-x')/v_g}{\tilde{t}}_p$, which shows $G^{(1)}(x',x;t)$ is trivially related to ${\tilde{t}}_p$. Thus, we can compute the coherent change in phase of the incident light from $G^{(1)}(x',x;t)$ at $x>0$ and $x'<0$. Below, we demonstrate that $G^{(1)}(x',x;t)$ can also be applied to extract the phase change and the Kerr effect for a multi-photon Fock state input.

We get the following expressions of $G^{(1)}(x',x;t)$ at $x>0$ and $x'<0$ for a coherent state input and a two-photon Fock state input with wave vector $k_p$:
\bea
G_c^{(1)}(x',x;t)&=&I_{cp} e^{i\om_p(x-x')/v_g}\f{i\Delta_p}{i\Delta_p-2\Gamma_p}\nn\\&\times&\Big(1-\f{8\Gamma_p^2v_gI_{cp}}{i\Delta_p(\Delta_p^2+4\Gamma_p^2)}+\mathcal{O}(\Omega_p^4)\Big),\label{KerrC}\\
G_2^{(1)}(x',x;t)&=&I_{2p} e^{i\om_p(x-x')/v_g}\f{i\Delta_p}{i\Delta_p-2\Gamma_p}\nn\\&\times&\Big(1-\f{8\Gamma_p^2v_gI_{1p}}{i\Delta_p(\Delta_p^2+4\Gamma_p^2)}+\f{2\Gamma_p v_gI_{1p}}{\Delta_p^2+4\Gamma_p^2}\Big),\label{KerrF}
%&&\f{G_c^{(1)}(x',x;t)}{I_c e^{i\om_p\f{(x-x')}{v_g}}}=\f{i\Delta}{2\Gamma+i\Delta}\Big(1+\f{8\Gamma^2v_gI_c}{i\Delta(4\Gamma^2+\Delta^2+2\Omega_p^2)}\Big),\\
%&&\f{G_2^{(1)}(x',x;t)}{I_1 e^{i\om_p\f{(x-x')}{v_g}}}=\f{i\Delta}{2\Gamma+i\Delta}\Big(1+\f{8\Gamma^2v_gI_1}{i\Delta(4\Gamma^2+\Delta^2)}+\f{2\Gamma v_gI_1}{4\Gamma^2+\Delta^2}\Big)
\eea
respectively.The expansion in order of $\Omega_p^2$ in Eq.~\ref{KerrC} is performed for a weak coherent state input, i.e., $\Omega_p^2/(2\Gamma_p^2)\ll1$. Here, $i\Delta_p/(i\Delta_p-2\Gamma_p) \equiv t_{1p}$ is the transmission amplitude of a single-photon or a faint coherent state input in the limit $\Omega_p \to 0$. Thus, the linear change in phase $\phi_p^{(1)}={\rm tan}^{-1}(-2\Gamma_p/\Delta_p)$.  The nonlinear phase change $\phi_p^{(2)}$ is obtained by the argument of the terms within the round brackets in Eqs.~\ref{KerrC},\ref{KerrF}. %\footnote{The form of the formulas in Eqs.~\ref{KerrC},\ref{KerrF} indicates the additive nature of linear and nonlinear contributions to the change in phase.}. 
The second term within the round brackets in Eq.~\ref{KerrC} for a coherent state matches that in Eq.~\ref{KerrF} for Fock state input if we replace $I_{cp}$ by $I_{1p}$. The appearance of $I_{1p}$ in Eq.~\ref{KerrF} is comprehensible since a single photon generates the nonlinear phase shift to the another for a two-photon input, and it also signals the nonlinear phase shift is smaller than the linear one by a factor of $1/\mathcal{L}$. 

Nevertheless, the third term in Eq.~\ref{KerrF} does not have an equivalent contribution in Eq.~\ref{KerrC}, and the term is due to the intermediate amplitudes of the excited emitter with one photon in the waveguide (see Eq.~\ref{FCer} in \cite{sm}). Thus, we find that $\phi_p^{(2)}$ for a two-photon Fock state input matches to that of a faint coherent state input (along with a replacement of $I_{cp}$ by $I_{1p}$) if we ignore the third term within the round brackets in Eq.~\ref{KerrF}. The third term in Eq.~\ref{KerrF} has a relatively small contribution to $\phi_p^{(2)}$ in comparison to the second term for the parameters of validity of the expression. From Eq.~\ref{KerrC}, we find the Kerr coefficient as $K=(1/E_p^2){\rm tan}^{-1}(8\Gamma_p^2 v_g I_{cp}/(\Delta_p(\Delta_p^2+\Gamma_p^2)))\approx 8\Gamma_p^2/(v_g\Delta_p(\Delta_p^2+\Gamma_p^2))$ for $\Omega_p^2/(2\Gamma_p^2)\ll1$ and $|\Delta_p|\geq\Gamma_p$. 

\begin{figure}
\includegraphics[width=0.99\linewidth]{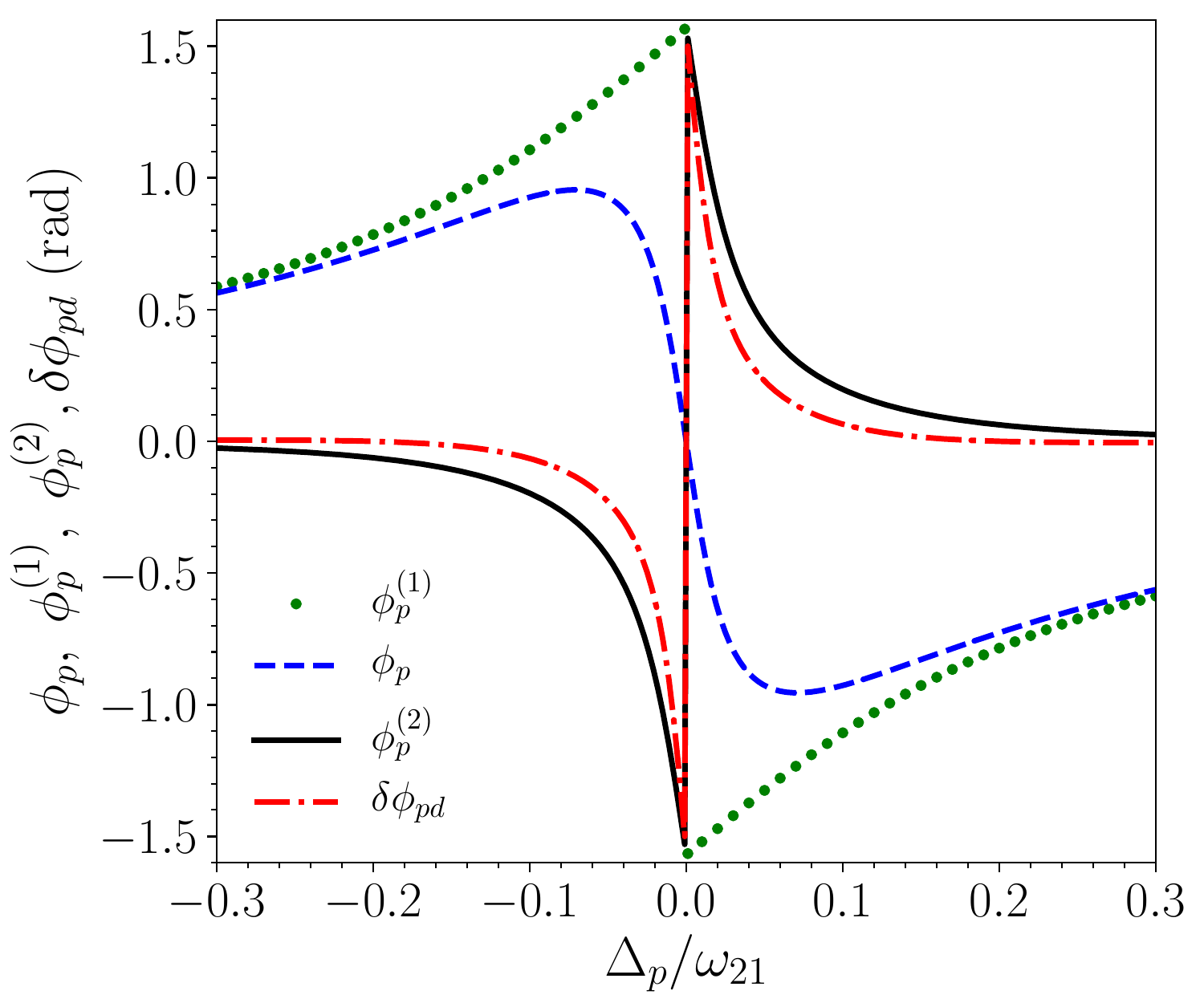}
\caption{Kerr versus cross-Kerr effect with Fock state photons in waveguide QED. The linear, nonlinear (Kerr) and total phase shifts $(\phi_p^{(1)}, \phi_p^{(2)}, \phi_p)$ of transmitted two photons inside an open waveguide side-coupled to a 2LE. The cross-Kerr phase shift $\delta \phi_{pd}$ of a single-photon probe beam by a single-photon drive beam, and both beams interact with two allowed transitions of a ladder-type 3LE.
 The parameters are $v_g=1,\Gamma_p/\omega_{21}=\Gamma_d/\omega_{21}=0.1, \Delta_d=0$ and $I_{1p}=I_{1d}=I_{cp}=0.0125 \omega_{21}/v_g.$}
\label{Kerrfig}
\end{figure}

A side-coupled 2LE perfectly reflects a resonant single-photon input, and the transmission phase shift $\phi_p^{(1)}$ is then not defined. However, two resonant photons can not be simultaneously perfectly reflected by a single emitter. Therefore, there will be a finite transmission for a two-photon resonant input pulse, and total phase shift $\phi_p$ will be zero at $\Delta_p=0$ due to the photon which passes the emitter without interacting with it. At $\Delta_p=0\pm$, we find $\phi_p^{(1)}=\mp \pi/2$ from $\phi_p^{(1)}={\rm tan}^{-1}(-2\Gamma_p/\Delta_p)$; thus we get $\phi_p^{(2)}=\pm \pi/2$ as $\phi_p \approx 0$. Magnitude of both $\phi_p^{(1)}$ and $\phi_p^{(2)}$ falls with increasing detuning $|\Delta_p|$. The linearity of $\phi_p^{(2)}$ with $E_p^2$ is true for small $\phi_p^{(2)}$ at $|\Delta_p| \ge \Gamma_p$. We show the above features of $\phi_p^{(1)}$ and $\phi_p^{(2)}$ with probe detuning $\Delta_p$ for a two-photon Fock state input in Fig.~\ref{Kerrfig}. 

{\it Cross-Kerr effect by 3LE:} Next, we consider a ladder-type 3LE with levels $|1\ra,|2\ra,|3\ra$. The energy difference between the excited states is $\hbar \om_{32}$. Two allowed optical transitions between the levels $|1\ra \leftrightarrow |2\ra$ and $|2\ra \leftrightarrow |3\ra$ are side-coupled to a probe and a drive beam of frequency $\om_p$ and $\om_d$, respectively. The full Hamiltonian of 3LE, light beams and their couplings reads:
\bea
\f{\mathcal{H}_3^k}{\hbar}&=&\om_{21} \sigma^{\dg}\sigma+ (\om_{32}+\om_{21})\mu^{\dg}\mu+\sum_k \Big(v_gk\sum_{\alpha=\pm}(a_{k\alpha}^{\dg}a_{k\alpha}\nn\\&-&b_{k\alpha}^{\dg}b_{k\alpha})+ g_p(\sigma^{\dg}\beta_{k+}+\beta_{k+}^{\dg}\sigma)+g_d(\mu^{\dg}\beta_{k-} +\beta_{k-}^{\dg}\mu) \Big),\nn\\\label{Ham3}
\eea
where we define $\beta_{k\pm}=(a_{k\pm}+b_{k\pm})$, and $\mu^{\dg}\equiv |3\ra \la 2|, \mu \equiv |2\ra \la 3|, \nu^{\dg} \equiv |1\ra \la 3|$ and $\nu \equiv |3\ra \la 1|$. Here, $a_{k\alpha}^{\dg}~[b_{k\alpha}^{\dg}]$  are creation operators for two different polarizations of right-moving [left-moving] photon modes of the probe and drive beams. The polarizations are denoted by subscript $\alpha=\pm$, and we choose $+$ and $-$ polarization, respectively, for the probe and drive beam. $g_p$ and $g_d$ are the respective coupling strength of the probe and drive beam with the 3LE. We assume that both the probe and drive input beams are incoming from the left of the 3LE.

For both the probe and drive input beams in the coherent states, the initial state at time $t_0$ is $|\psi\ra=|E_p,\om_p\ra \otimes |E_d,\om_d\ra$, which satisfies $a_{k+}(t_0)|\psi \ra=(\sqrt{\mathcal{L}}E_p/v_g) \delta_{k,\om_p/v_g}|\psi \ra$, $a_{k-}(t_0)|\psi \ra=(\sqrt{\mathcal{L}}E_d/v_g) \delta_{k,\om_d/v_g}|\psi\ra$, where $E_p$ and $E_d$ are their respective (real) amplitude. Further, the intensity of probe and drive beam are $I_{cp}=E_p^2/v_g^2$ and $I_{cd}=E_d^2/v_g^2$. For Fock state input, we consider the probe and drive beams consisting of single photons as 
\bea
|k_p,k_d\ra=\int_{-\mathcal{L}/2}^{\mathcal{L}/2}\int_{-\mathcal{L}/2}^{\mathcal{L}/2}\f{dx_1dx_2}{\mathcal{L}} e^{i(k_px_1+k_dx_2)}\tilde{a}_{x_1+}^{\dg}\tilde{a}_{x_2-}^{\dg}|\varphi\ra,\nn
\eea
where $\tilde{a}_{x\pm}(t)=\sum_k e^{ikx}a_{k\pm}(t)/\sqrt{\mathcal{L}}$ and $\om_d=v_gk_d$. The intensity of the single-photon probe and drive beam are respectively $I_{1p}=1/\mathcal{L}$ and $I_{1d}=1/\mathcal{L}$. %To find the cross-Kerr effect in the system, we derive the outgoing scattered states of the probe and drive beams within the Schr{\"o}dinger picture for the Fock state inputs, and calculate the time-evolution of operators of the full system within the Heisenberg picture for the coherent state inputs.

A ladder-type 3LE made of a superconducting artificial atom was used to demonstrate an effective interaction between two different light beams at the single-photon quantum regime in Ref.~\cite{HoiPRL2013}. Such an effective interaction is essentially similar in physical mechanism to the above explored effective interaction between photons of a single beam induced by the Kerr nonlinearity of the medium (e.g., an emitter). This effective coupling between multiple beams is known as the cross-Kerr effect. The photon-photon interaction in a cross-Kerr medium has been utilized to propose quantum nondemolition measurement of a single propagating microwave photon with high fidelity \cite{Sathyamoorthy2014}. Extending the earlier discussion of the optical Kerr effect, the cross-Kerr effect can be interpreted as modulation of refractive index or a phase change of a probe beam due to a drive beam. Thus, we write the total change in phase of the probe beam in a cross-Kerr medium as  $\phi_{pd}=\phi_{p}^{(1)}+\delta\phi_{pd}$, where $\phi_{p}^{(1)}$ again indicates the linear change in phase of the probe beam in the absence of drive beam, and $\delta \phi_{pd}$ captures the change in phase of the probe beam in the presence of the drive beam. In analogy to the Kerr coefficient, we define the cross-Kerr coefficient $K_c$ using $\delta \phi_{pd}\equiv \phi_{pd}|_{\Omega_d\ne 0}-\phi_{pd}|_{\Omega_d=0}=K_cE_d^2$, where $\Omega_d=\bar{g}_dE_d/v_g$ is the Rabi frequency of the coherent drive beam and $\bar{g}_d=\sqrt{\mathcal{L}}g_d$. We can find $\phi_{pd}$ of a probe beam in the presence and absence of a drive beam using $G_+^{(1)}(x',x;t)=\la \tilde{a}^{\dg}_{x'+}(t)\tilde{a}_{x+}(t)\ra$ at $x>0$ and $x'<0$.

To find the cross-Kerr effect in our system, we derive the outgoing scattered states of the probe and drive beams within the Schr{\"o}dinger picture for the Fock state inputs and calculate the time-evolution of operators of the full system within the Heisenberg picture for the coherent state inputs. The first-order coherence $G_+^{(1)}(x',x;t)$ at $x>0$ and $x'<0$ for coherent state and Fock state inputs are
\bea
&&G_{+c}^{(1)}(x',x;t)=I_{cp} e^{i\om_p(x-x')/v_g}\f{i\Delta_p}{i\Delta_p-2\Gamma_p}\nn\\&\times&\Big(1-\f{4\Gamma_p\Gamma_d v_gI_{cd}}{i\Delta_p(i\Delta_p-2\Gamma_p)(i(\Delta_p+\Delta_d)-2\Gamma_d)}+\mathcal{O}(\Omega_d^4) \Big),\nn\\\label{crKerrC}\\
&&G_{+2}^{(1)}(x',x;t)=I_{1p} e^{i\om_p(x-x')/v_g}\f{i\Delta_p}{i\Delta_p-2\Gamma_p}\nn\\&\times&\Big(1-\f{4\Gamma_p\Gamma_d v_gI_{1d}}{i\Delta_p(i\Delta_p-2\Gamma_p)(i(\Delta_p+\Delta_d)-2\Gamma_d)}\Big),\label{crKerrF}
%&&\f{G_c^{(1)}(x',x;t)}{I_c e^{i\om_p\f{(x-x')}{v_g}}}=\f{i\Delta}{2\Gamma+i\Delta}\Big(1+\f{8\Gamma^2v_gI_c}{i\Delta(4\Gamma^2+\Delta^2+2\Omega_p^2)}\Big),\\
%&&\f{G_2^{(1)}(x',x;t)}{I_1 e^{i\om_p\f{(x-x')}{v_g}}}=\f{i\Delta}{2\Gamma+i\Delta}\Big(1+\f{8\Gamma^2v_gI_1}{i\Delta(4\Gamma^2+\Delta^2)}+\f{2\Gamma v_gI_1}{4\Gamma^2+\Delta^2}\Big)
\eea
where we have expanded Eq.~\ref{crKerrC} in order of $\Omega_d^2$ for a faint coherent state drive beam. Here, $\Delta_d=\om_d-\om_{32}, \Gamma_d=\bar{g}_d^2/(2v_g)$. We find from Eqs.~\ref{crKerrC},\ref{crKerrF} that the leading order contribution from a faint coherent state drive beam to $\delta\phi_{pd}$ matches that from a single-photon drive when we identify $I_{cd}$ by $I_{1d}$. While the cross-Kerr phase shift $\delta\phi_{pd}$ depends on both probe and drive photon detuning, let us study it for $\Delta_d=0$ when it shows a relatively large value. The features of $\delta\phi_{pd}$ with $\Delta_p$ is similar to $\phi_p^{(2)}$ for single photons as shown in Fig.~\ref{Kerrfig}. However, the value of $\delta\phi_{pd}$ at $\Delta_d=0$ in Fig.~\ref{Kerrfig} is always smaller than $\phi_p^{(2)}$ at any finite $\Delta_p$ for single photons \cite{sm}. Therefore, the Kerr nonlinearity $K$ by a 2LE between two single photons is relatively higher than the cross-Kerr nonlinearity $K_c$ by a ladder-type 3LE between them. Nevertheless, the value of $K_c$ depends on the type of 3LE, which is determined by the optical transitions used for the drive and probe beams \cite{Vinu2020}. For example, we find $K_c$ for a single-photon probe, and drive beam can be higher for a $V$-type 3LE than a ladder-type 3LE. Further, $K_c$ by a $V$-type 3LE can be slightly higher than $K$ by a 2LE between two Fock state photons.  

Our findings for comparing various linear and nonlinear light scattering by a single emitter embedded in an open waveguide for different light sources will benefit the rapid progress of waveguide QED. Experiments with superconducting circuits can potentially verify our theoretical predictions in this work. In future studies, we hope to compare different light sources for more complex waveguide QED setups, such as with giant atoms, separated emitters, and topological waveguides.

{\it Acknowledgments}
We acknowledge funding from the Ministry of Electronics $\&$ Information Technology (MeitY), India under the grant for ``Centre for Excellence in Quantum Technologies'' with Ref. No. 4(7)/2020-ITEA.

\bibliography{bibliography2}

\clearpage
  \widetext
    \begin{center}
    \textbf{\large Supplemental Materials for\\`Single photons versus coherent state input in waveguide quantum electrodynamics: light scattering, Kerr and cross-Kerr effect'} \\ \vspace{0.3cm}
  Athul Vinu and Dibyendu Roy
    \end{center}
    \setcounter{equation}{0}
    \setcounter{figure}{0}
    \setcounter{table}{0}
    \setcounter{page}{1}
    \setcounter{section}{0}
    \makeatletter
    \renewcommand{\thesection}{S-\Roman{section}}
    \renewcommand{\theequation}{S\arabic{equation}}
    \renewcommand{\thefigure}{S\arabic{figure}}
     \renewcommand{\thetable}{S\arabic{table}}
    \renewcommand{\bibnumfmt}[1]{[S#1]}
    \renewcommand{\citenumfont}[1]{S#1}
    %\end{widetext}
 
\section{Scattering of single- and two-photon Fock states by a two-level emitter}

The real-space Hamiltonian of a two-level emitter (2LE) side-coupled to a linear waveguide with right-moving and left-moving photon modes:
\bea
\f{\mathcal{H}^x_2}{\hbar}=\om_{21} \sigma^{\dg}\sigma-iv_g\int_{-\mathcal{L}/2}^{\mathcal{L}/2}dx \big(\tilde{a}^{\dg}_x\partial_x\tilde{a}_x-\tilde{b}_x^{\dg}\partial_x\tilde{b}_x\big)+\bar{g}_p\sigma^{\dg}(\tilde{a}_{0}+\tilde{b}_{0})+\bar{g}_p(\tilde{a}_{0}^{\dg}+\tilde{b}_{0}^{\dg})\sigma.\label{Ham2rs}
\eea
First, we consider a single-photon input state with wave vector $k_p$ and frequency $\om_{p}=v_gk_p$ from the left of the emitter:
\bea
|k_p\ra=\f{1}{\sqrt{\mathcal{L}}}\int_{-\mathcal{L}/2}^{\mathcal{L}/2}dx\:e^{ik_px}\tilde{a}_{x}^{\dg}|\varphi\ra,\label{1phIS}
\eea
which satisfies $I_{1p}=\la k_p|\int dk\:a^{\dg}_{k}a_{k}|k_p\ra/\mathcal{L}=1/\mathcal{L}$. The full single-photon state $|k_p^+\ra$ of the Hamiltonian is derived from $\mathcal{H}^x_2|k_p^+\ra=\hbar v_g k_p|k_p^+\ra$ using the initial conditions in Eq.~\ref{1phIS}. It is given as
\bea
|k_p^+\ra=\int_{-\mathcal{L}/2}^{\mathcal{L}/2} dx\Big[g_R(x)\tilde{a}_{x}^{\dg}+g_L(x)\tilde{b}_{x}^{\dg}+\delta(x)\tilde{e}_p\sigma^{\dg}\Big]|\varphi\ra|g\ra,
\eea
whose the amplitudes are
\bea
g_R(x)=\f{e^{ik_px}}{\sqrt{\mathcal{L}}}(\theta(-x)+t_{1p}\theta(x)),~g_L(x)=\f{e^{-ik_px}}{\sqrt{\mathcal{L}}}r_{1p}\theta(-x),~e_p=\f{\bar{g}_p}{\Delta_p+2i\Gamma_p},~t_{1p}=\f{\Delta_p}{\Delta_p+2i\Gamma_p}.
\eea
Here, $t_{1p}$ and $r_{1p}=t_{1p}-1$ are, respectively, single-photon transmission and reflection amplitude, and the amplitude of emitter's excitation is $\tilde{e}_p=e_p/\sqrt{\mathcal{L}}$.

The normalized two-photon incident Fock state with degenerate wave vectors $\bold{k}_p=(k_p,k_p)$ in the right-moving channels:
\bea
|\bm{k}_p\ra=\f{1}{\mathcal{L}}\int_{-\mathcal{L}/2}^{\mathcal{L}/2}\int_{-\mathcal{L}/2}^{\mathcal{L}/2}dx_1dx_2\:e^{ik_p(x_1+x_2)}\f{1}{\sqrt{2}}\tilde{a}_{x_1}^{\dg}\tilde{a}_{x_2}^{\dg}|\varphi\ra, \label{2phIS}
\eea
which has intensity $I_{2p}=2/\mathcal{L}$. The two-photon state of the Hamiltonian including the scattered and incident parts is 
\bea
|\bm{k}_p^+\ra&=&\int_{-\mathcal{L}/2}^{\mathcal{L}/2}\int_{-\mathcal{L}/2}^{\mathcal{L}/2} dx_1dx_2\Big[g_{RR}(x_1,x_2)\f{1}{\sqrt{2}}\tilde{a}_{x_1}^{\dg}\tilde{a}_{x_2}^{\dg}+g_{RL}(x_1,x_2)\tilde{a}_{x_1}^{\dg}\tilde{b}_{x_2}^{\dg}+g_{LL}(x_1,x_2)\f{1}{\sqrt{2}}\tilde{b}_{x_1}^{\dg}\tilde{b}_{x_2}^{\dg}\nn\\&&+\big(e_R(x_1)\tilde{a}_{x_1}^{\dg}+e_L(x_1)\tilde{b}_{x_1}^{\dg}\big)\delta(x_2)\sigma^{\dg}\Big]|\varphi\ra|g\ra,
\eea
whose amplitudes can be found by solving a set of linear, coupled, inhomogeneous different equations obtained from $\mathcal{H}^x_2|\bm{k}_p^+\ra=2\hbar v_g k_p|\bm{k}_p^+\ra$ with the initial conditions set by Eq.~\ref{2phIS}. These amplitudes are
\bea
g_{RR}(x_1,x_2)&=&g_R(x_1)g_R(x_2)+\Big[\f{2\Gamma_p}{v_g}\tilde{e}_{p}^2\:e^{i(2v_gk_p-\om_{21}+2i\Gamma_p)x_2/v_g}e^{i(\om_{21}-2i\Gamma_p)x_1/v_g}\theta(x_2-x_1)\theta(x_1)+(x_1 \leftrightarrow x_2)\Big], \nn\\
g_{RL}(x_1,x_2)&=&\sqrt{2}g_R(x_1)g_L(x_2)+\Big[\f{2\sqrt{2}\Gamma_p}{v_g}\tilde{e}_{p}^2\:e^{-i(2v_gk_p-\om_{21}+2i\Gamma_p)x_2/v_g}e^{i(\om_{21}-2i\Gamma_p)x_1/v_g}\theta(|x_2|-x_1)\theta(x_1)\theta(-x_2)+(x_1 \leftrightarrow x_2)\Big], \nn\\
g_{LL}(x_1,x_2)&=& g_L(x_1)g_L(x_2)+\Big[\f{2\Gamma_p}{v_g}\tilde{e}_{p}^2\:e^{-i(2v_gk_p-\om_{21}+2i\Gamma_p)x_2/v_g}e^{-i(\om_{21}-2i\Gamma_p)x_1/v_g}\theta(|x_2-x_1|)\theta(-x_1)\theta(-x_2)+(x_1 \leftrightarrow x_2)\Big],\nn\\
e_{R}(x)&=&\sqrt{2}\:g_R(x)\tilde{e}_p+\f{\sqrt{2}i\bar{g}_p}{v_g}\tilde{e}_p^2\:e^{i(2v_gk_p-\om_{21}+2i\Gamma_p)x/v_g}\theta(x),\nn\\
e_{L}(x)&=&\sqrt{2}\:g_L(x)\tilde{e}_p+\f{\sqrt{2}i\bar{g}_p}{v_g}\tilde{e}_p^2\:e^{-i(2v_gk_p-\om_{21}+2i\Gamma_p)x/v_g}\theta(-x),
\eea
where $\theta(x)$ is the Heaviside step function. 
\subsection{First-order coherence:}
Using the single-photon and two-photon states of the Hamiltonian, we can find the first-order coherence $G^{(1)}(x',x;t)$ of the transmitted photon(s) for incident photon(s) in the right-moving channel(s). We get for the single-photon case with $x>0,x'<0$:
\bea
G^{(1)}_1(x',x;t)=\la k_p^+|\tilde{a}^{\dg}_{x'}\tilde{a}_{x}|k_p^+\ra=I_{1p}t_{1p}e^{ik_p(x-x')}\equiv I_{1p}\f{i\Delta_p}{i\Delta_p-2\Gamma_p}e^{ik_p(x-x')},
\eea
which gives the linear change in phase $\phi_p^{(1)}={\rm tan}^{-1}(-2\Gamma_p/\Delta_p)$. For two photons, we find again for $x>0,x'<0$
\bea
G^{(1)}_2(x',x;t)=\la \bm{k}_p^+|\tilde{a}^{\dg}_{x'}\tilde{a}_{x}|\bm{k}_p^+\ra=\int_{-\mathcal{L}/2}^{\mathcal{L}/2}dy\big[2g^*_{RR}(x',y)g_{RR}(x,y)+g^*_{RL}(x',y)g_{RL}(x,y)\big]+e^*_R(x')e_R(x).
\eea 
Below we give each term in the above relation explicitly to show how to get the final result. We find 
\bea
&&2\int_{-\mathcal{L}/2}^{\mathcal{L}/2}dy\:g^*_{RR}(x',y)g_{RR}(x,y)\nn\\
&&=\f{1}{\mathcal{L}}e^{ik_p(x-x')}\Big[\big(1+|t_{1p}|^2\big)t_{1p}-\f{4}{\mathcal{L}}r_{1p}t^*_{1p}e_p^2+\f{2}{\mathcal{L}} r_{1p}t^*_{1p}e_p^2\Big(e^{i(\Delta_p+2i\Gamma_p)x/v_g}+e^{-i(\Delta_p+2i\Gamma_p)(x-\mathcal{L}/2)/v_g}\Big)\Big],\label{FCgrr}\\
&&\int_{-\mathcal{L}/2}^{\mathcal{L}/2}dy\:g^*_{RL}(x',y)g_{RL}(x,y)\nn\\
&&=\f{1}{\mathcal{L}}e^{ik_p(x-x')}\Big[|r_{1p}|^2t_{1p}-\f{2}{\mathcal{L}}|r_{1p}|^2e_p^2\Big(1-e^{-i(\Delta_p+2i\Gamma_p)(x-\mathcal{L}/2)/v_g}\Big)+\f{2}{\mathcal{L}}|r_{1p}|^2e_p^2\Big(e^{i(\Delta_p+2i\Gamma_p)x/v_g}-1\Big)\Big], \label{FCgrl}\\
&&e^*_R(x')e_R(x)=\f{2}{\mathcal{L}^2}|e_p|^2e^{ik_p(x-x')}\Big[t_{1p}-r_{1p}e^{i(\Delta_p+2i\Gamma_p)x/v_g}\Big].\label{FCer}
\eea
We add Eqs.~\ref{FCgrr}-\ref{FCer} to find that the terms with a factor $e^{i(\Delta_p+2i\Gamma_p)x/v_g}$ disappear. The terms with a factor $e^{-i(\Delta_p+2i\Gamma_p)(x-\mathcal{L}/2)/v_g}$ also vanish when $\mathcal{L}>>v_g/\Gamma_p$. Thus, we finally get $G^{(1)}_2(x',x;t)$ as in the main text.
\bea
G^{(1)}_2(x',x;t)&=&\f{1}{\mathcal{L}}e^{ik_p(x-x')}\Big[\big(1+|t_{1p}|^2+|r_{1p}|^2\big)t_{1p}-\f{4}{\mathcal{L}}r_{1p}t^*_{1p}e_p^2-\f{4}{\mathcal{L}}|r_{1p}|^2e_p^2+\f{2}{\mathcal{L}}|e_p|^2t_{1p}\Big]\nn\\
&=&I_{2p}\f{i\Delta_p}{i\Delta_p-2\Gamma_p}e^{ik_p(x-x')}\Big[1-\f{8\Gamma_p^2v_gI_{1p}}{i\Delta_p(\Delta_p^2+4\Gamma_p^2)}+\f{2\Gamma_p v_gI_{1p}}{\Delta_p^2+4\Gamma_p^2}\Big]. \label{FCerF}
\eea
\subsection{Reflection current}
We evaluate the expectation of reflection current operator in Eq.~\ref{refcur} in the full single-photon and two-photon states. We remind that the contributions in reflection current arise solely from scattered photons by the emitter. We find
\bea
\mathcal{J}_1&=&i\bar{g}_p\la k_p^+|(\sigma^{\dg}\tilde{b}_0- \tilde{b}^{\dg}_0\sigma)|k_p^+\ra = \f{v_g}{\mathcal{L}}|r_{1p}|^2, \nn\\
\mathcal{J}_2&=&i\bar{g}_p\la \bm{k}_p^+|(\sigma^{\dg}\tilde{b}_0- \tilde{b}^{\dg}_0\sigma)|\bm{k}_p^+\ra =2{\rm Re}\Big[ i\bar{g}_p \int_{-\mathcal{L}/2}^{\mathcal{L}/2}dx~(e^*_R(x)g_{RL}(x,0)+\sqrt{2}e^*_L(x)g_{LL}(x,0))\Big]. \label{refcur2p}
\eea
Each part of the two-photon reflection current is given as following.
\bea
 i\bar{g}_p \int_{-\mathcal{L}/2}^{\mathcal{L}/2}dx e^*_R(x)g_{RL}(x,0)&=&\f{i\bar{g}_p}{2\mathcal{L}}e_p^*r_{1p}+\f{\Gamma_p}{\mathcal{L}}|t_{1p}|^2|e_p|^2 +\f{\Gamma_p}{\mathcal{L}^2}|e_p|^4(1-e^{-2\Gamma_p \mathcal{L}/v_g})\nn\\
 &+&\f{2\Gamma_p}{\mathcal{L}^2}\Big[e_p^3e_p^*t_{1p}^*\Big(e^{i(\Delta_p+2i\Gamma_p)\mathcal{L}/(2v_g)}-1\Big)+e_p^{*3}e_pt_{1p}\Big(e^{i(-\Delta_p+2i\Gamma_p)\mathcal{L}/(2v_g)}-1\Big)\Big], \label{2p1}\\
\sqrt{2}e^*_L(x)g_{LL}(x,0))&=&\f{i\bar{g}_p}{2\mathcal{L}}|r_{1p}|^2e_p^*r_{1p}+\f{\Gamma_p}{\mathcal{L}^2}|e_p|^4(1-e^{-2\Gamma_p \mathcal{L}/v_g})+\f{2\Gamma_p v_g}{i\mathcal{L}^2\bar{g}_p}\Big[r_{1p}^2e_p^{*3}\Big(1-e^{i(-\Delta_p+2i\Gamma_p)\mathcal{L}/(2v_g)}\Big)\nn\\
&-&r_{1p}^{*2}e_p^{3}\Big(1-e^{i(\Delta_p+2i\Gamma_p)\mathcal{L}/(2v_g)}\Big)\Big].\label{2p2}
\eea
In the limit of $\mathcal{L}>>v_g/\Gamma_p$, we find the following by adding Eqs.~\ref{2p1}, \ref{2p2}:
\bea
\mathcal{J}_2&=&\f{2v_g}{\mathcal{L}}|r_{1p}|^2-\f{16v_g^2\Gamma_p^3}{\mathcal{L}^2(\Delta_p^2+4\Gamma_p^2)^2}=\f{2}{\mathcal{L}}\f{4v_g\Gamma_p^2}{\Delta_p^2+4\Gamma_p^2}-\f{1}{\mathcal{L}^2}\f{16v_g^2\Gamma_p^3}{(\Delta_p^2+4\Gamma_p^2)^2}.
\eea

\section{Scattering of probe and drive single-photon Fock states by a ladder-type three-level emitter}
We consider a ladder-type three-level emitter (3LE) with two allowed transitions being side-coupled to a linear waveguide carrying right-moving and left-moving photon modes of a probe and a drive beam. The real-space Hamiltonian of Eq.~\ref{Ham3} is given by
\bea
\f{\mathcal{H}^x_3}{\hbar}&=&\om_{21} \sigma^{\dg}\sigma+(\om_{32}+\om_{21})\mu^{\dg}\mu-iv_g\int_{-\mathcal{L}/2}^{\mathcal{L}/2}dx\sum_{\alpha=\pm}\big(\tilde{a}^{\dg}_{x\alpha}\partial_x\tilde{a}_{x\alpha}-\tilde{b}_{x\alpha}^{\dg}\partial_x\tilde{b}_{x\alpha}\big)+\bar{g}_p\sigma^{\dg}(\tilde{a}_{0+}+\tilde{b}_{0+})+\bar{g}_p(\tilde{a}_{0+}^{\dg}+\tilde{b}_{0+}^{\dg})\sigma\nn \\&+&\bar{g}_d\mu^{\dg}(\tilde{a}_{0-}+\tilde{b}_{0-})+\bar{g}_d(\tilde{a}_{0-}^{\dg}+\tilde{b}_{0-}^{\dg})\mu, \label{Ham3rs}
\eea
where $\bar{g}_d=\sqrt{\mathcal{L}}g_d$.  Here, $\tilde{a}_{x\alpha}^{\dg}(t)=\int dk\:e^{-ikx}a_{k\alpha}^{\dg}(t)/\sqrt{\mathcal{L}}$ and $\tilde{b}_{x\alpha}^{\dg}(t)=\int dk\:e^{-ikx}b_{k\alpha}^{\dg}(t)/\sqrt{\mathcal{L}}$ are creation operators at position $x \in [-\mathcal{L}/2,\mathcal{L}/2]$ of right-moving and left-moving photon modes of the probe $(\alpha=+)$ and drive $(\alpha=-)$ beams. %To obtain a real-space description of the propagating photons at position $x \in [-\mathcal{L}/2,\mathcal{L}/2]$, we define $\tilde{a}_{x}(t)=\int dk\:e^{ikx}a_{k}(t)/\sqrt{\mathcal{L}}$ and $\tilde{b}_{x}(t)=\int dk\:e^{ikx}b_{k}(t)/\sqrt{\mathcal{L}}$. 

The input state of a single-photon probe and a single-photon drive beam with respective wave vector $k_p$ and $k_d$ (with corresponding frequencies $\om_{p}=v_gk_p$ and $\om_{d}=v_gk_d$) in the right-moving channel is:
\bea
|k_p,k_d\ra=\f{1}{\mathcal{L}}\int_{-\mathcal{L}/2}^{\mathcal{L}/2}\int_{-\mathcal{L}/2}^{\mathcal{L}/2}dx_1dx_2\:e^{i(k_p x_1+k_d x_2)}\tilde{a}_{x_1+}^{\dg}\tilde{a}_{x_2-}^{\dg}|\varphi\ra,\label{phIS}
\eea
which satisfies $I_{1p}=\la k_p,k_d|\int dk\:a^{\dg}_{k+}a_{k+}|k_p,k_d\ra/\mathcal{L}=1/\mathcal{L}$ and $I_{1d}=\la k_p,k_d|\int dk\:a^{\dg}_{k-}a_{k-}|k_p,k_d\ra/\mathcal{L}=1/\mathcal{L}$.

The full two-photon state of the Hamiltonian $\mathcal{H}^x_3$ including the scattered and incident photons is 
\bea
|k_p,k_d^+\ra&=&\int_{-\mathcal{L}/2}^{\mathcal{L}/2}\int_{-\mathcal{L}/2}^{\mathcal{L}/2} dx_1dx_2\Big[\tilde{g}_{RR}(x_1,x_2)\tilde{a}_{x_1+}^{\dg}\tilde{a}_{x_2-}^{\dg}+\tilde{g}_{RL}(x_1,x_2)\tilde{a}_{x_1+}^{\dg}\tilde{b}_{x_2-}^{\dg}+\tilde{g}_{LR}(x_1,x_2)\tilde{b}_{x_1+}^{\dg}\tilde{a}_{x_2-}^{\dg}\nn\\&&+\tilde{g}_{LL}(x_1,x_2)\tilde{b}_{x_1+}^{\dg}\tilde{b}_{x_2-}^{\dg}+\big(\tilde{e}_R(x_2)\tilde{a}_{x_2-}^{\dg}+\tilde{e}_L(x_2)\tilde{b}_{x_2-}^{\dg}\big)\delta(x_1)\sigma^{\dg}\Big]|\varphi\ra|g\ra,
\eea
whose amplitudes can be found by solving a set of linear, coupled, inhomogeneous different equations obtained from $\mathcal{H}^x_3|k_p,k_d^+\ra=\hbar v_g (k_p+k_d)|k_p,k_d^+\ra$ with the initial conditions set by Eq.~\ref{phIS}. These amplitudes are
\bea
\tilde{g}_{RR}(x_1,x_2)&=&g_R(x_1)\f{e^{ik_dx_2}}{\sqrt{\mathcal{L}}}-\Big[\f{2\sqrt{\Gamma_p\Gamma_d}}{v_g}\tilde{e}_{p}\tilde{e}_{d}\:e^{i(k_px_1+k_dx_2)}e^{i(\Delta_{p}+2i\Gamma_p)(x_2-x_1)/v_g}\theta(x_2-x_1)\theta(x_1)\Big], \nn\\
\tilde{g}_{LR}(x_1,x_2)&=&g_L(x_1)\f{e^{ik_dx_2}}{\sqrt{\mathcal{L}}}-\Big[\f{2\sqrt{\Gamma_p\Gamma_d}}{v_g}\tilde{e}_{p}\tilde{e}_{d}\:e^{i(-k_px_1+k_dx_2)}e^{i(\Delta_{p}+2i\Gamma_p)(x_2+x_1)/v_g}\theta(x_2+x_1)\theta(-x_1)\Big], \nn\\
\tilde{g}_{RL}(x_1,x_2)&=&-\f{2\sqrt{\Gamma_p\Gamma_d}}{v_g}\tilde{e}_{p}\tilde{e}_{d}\:e^{i(k_px_1-k_dx_2)}e^{-i(\Delta_{p}+2i\Gamma_p)(x_2+x_1)/v_g}\theta(-x_2-x_1)\theta(x_1), \nn\\
\tilde{g}_{LL}(x_1,x_2)&=& -\f{2\sqrt{\Gamma_p\Gamma_d}}{v_g}\tilde{e}_{p}\tilde{e}_{d}\:e^{-i(k_px_1+k_dx_2)}e^{-i(\Delta_{p}+2i\Gamma_p)(x_2-x_1)/v_g}\theta(-x_2+x_1)\theta(-x_1),\nn\\
\tilde{e}_{R}(x)&=&\f{e^{ik_dx}}{\sqrt{\mathcal{L}}}\tilde{e}_p-\f{i\bar{g}_d}{v_g}\tilde{e}_p\tilde{e}_{d}\:e^{i(v_g(k_p+k_d)-\om_{21}+2i\Gamma_p)x/v_g}\theta(x),\nn\\
\tilde{e}_{L}(x)&=&-\f{i\bar{g}_d}{v_g}\tilde{e}_p\tilde{e}_{d}\:e^{-i(v_g(k_p+k_d)-\om_{21}+2i\Gamma_p)x/v_g}\theta(-x),~~e_d=\f{\bar{g}_d}{\Delta_p+\Delta_d+2i\Gamma_d},
\eea
where $\Gamma_d=\bar{g}_d^2/(2v_g),~\Delta_d=\om_d-\om_{32}$ and $\tilde{e}_d=e_d/\sqrt{\mathcal{L}}$. 

Next, we calculate the first-order coherence $G^{(1)}_{+2}(x',x;t)$ of the transmitted probe photon for incident probe and drive photons in the right-moving channels. We find for $x>0,x'<0$
\bea
G^{(1)}_{+2}(x',x;t)=\la k_p,k_d^+|\tilde{a}^{\dg}_{x'+}\tilde{a}_{x+}|k_p,k_d^+\ra=\int_{-\mathcal{L}/2}^{\mathcal{L}/2}dy\big[\tilde{g}^*_{RR}(x',y)\tilde{g}_{RR}(x,y)+\tilde{g}^*_{RL}(x',y)\tilde{g}_{RL}(x,y)\big].
\eea 
We find $\int_{-\mathcal{L}/2}^{\mathcal{L}/2}dy\:\tilde{g}^*_{RL}(x',y)\tilde{g}_{RL}(x,y)=0$, and 
\bea
\int_{-\mathcal{L}/2}^{\mathcal{L}/2}dy\:\tilde{g}^*_{RR}(x',y)\tilde{g}_{RR}(x,y)=\f{1}{\mathcal{L}}e^{ik_p(x-x')}\Big[t_{1p}-\f{i\tilde{g}_d}{v_g}e_{d}\tilde{e}_p^2 \Big(1-e^{-i(\Delta_p + 2i\Gamma_p)(x-\mathcal{L}/2)/v_g}\Big)\Big]. \label{crossKSM1}
\eea
The term with a factor $e^{-i(\Delta_p+2i\Gamma_p)(x-\mathcal{L}/2)/v_g}$ in Eq.~\ref{crossKSM1} vanishes when $\mathcal{L}>>v_g/\Gamma_p$. Thus, we finally get $G^{(1)}_{+2}(x',x;t)$ as 
\bea
G^{(1)}_{+2}(x',x;t)=I_{1p} e^{i\om_p(x-x')/v_g}\f{i\Delta_p}{i\Delta_p-2\Gamma_p}\Big(1-\f{4\Gamma_p\Gamma_d v_gI_{1d}}{i\Delta_p(i\Delta_p-2\Gamma_p)(i(\Delta_p+\Delta_d)-2\Gamma_d)}\Big).\label{crossKSM2}
\eea 
Next, we compare the cross-Kerr phase shift $\delta \phi_{pd}$ of a probe photon due to a drive photon with the Kerr phase shift $\phi_p^{(2)}$ between two Fock state photons. For this,  we write Eq.~\ref{crossKSM2} in the limit of $\Delta_d=0$, $\Gamma_d=\Gamma_p$ and $I_{1d}=I_{1p}$:  
\bea
G^{(1)}_{+2}(x',x;t)=I_{1p} e^{i\om_p(x-x')/v_g}\f{i\Delta_p}{i\Delta_p-2\Gamma_p}\Big(1-\f{4\Gamma_p^2 v_gI_{1p}}{i\Delta_p(i\Delta_p-2\Gamma_p)^2}\Big).\label{crossKSM3}
\eea 
The second term within the round brackets in Eq.~\ref{crossKSM3} is a bit similar to the second term within the round brackets in Eq.~\ref{FCerF} except a half factor difference in the numerator and the appearance of $(i\Delta_p-2\Gamma_p)^2$ instead of $|i\Delta_p-2\Gamma_p|^2$ in the denominator.
\section{Scattering of a coherent state input by a two-level emitter}
The momentum-space Hamiltonian of a 2LE side-coupled to a waveguide with linearized energy-momentum dispersion of photons is
\bea
\f{\mathcal{H}^k_2}{\hbar}&=&\om_{21} \sigma^{\dg}\sigma+ \sum_{k}\Big(v_gk \big(a_{k}^{\dg}a_{k}-b_{k}^{\dg}b_{k}\big)+g_p\sigma^{\dg}(a_{k}+b_{k})+g_p(a_{k}^{\dg}+b_{k}^{\dg})\sigma \Big).
\eea
We first write the Heisenberg equations for different operators appearing in the above Hamiltonian. Then, we formally solve these equations for the photon field operators as
\bea
a_k(t)&=&e^{-iv_gkt}a_k(t_0)-ig_p\int_{t_0}^tdt' e^{-iv_gk(t-t')}\sigma(t'), \label{af}\\
b_k(t)&=&e^{iv_gkt}b_k(t_0)-ig_p\int_{t_0}^tdt' e^{iv_gk(t-t')}\sigma(t'), \label{bf}
\eea
where $a_k(t_0)$ and $b_k(t_0)$ are initial photon fields at $t=t_0$ before the fields  interacting with the 2LE. We plug $a_k(t)$ and $b_k(t)$ in Eqs.~\ref{af}, \ref{bf} in the Heisenberg equations of the emitter operators and rewrite these equations as
\bea
\f{d\sigma(t)}{dt}&=&(-i\omega_{21}-2\Gamma_p)\sigma(t)-ig_p(1-2\sigma^{\dg}(t)\sigma(t))(\eta_a(t)+\eta_b(t)), \label{sig1}\\
\f{d\sigma^{\dg}(t)\sigma(t)}{dt}&=&-4\Gamma_p\sigma^{\dg}(t)\sigma(t)-ig_p\sigma^{\dg}(t)(\eta_a(t)+\eta_b(t))+ig_p(\eta_a^{\dg}(t)+\eta_b^{\dg}(t))\sigma(t), \label{sigcor1}
\eea 
where $\eta_a(t)=\sum_{k}e^{-iv_gk(t-t_0)}a_k(t_0)$ and $\eta_b(t)=\sum_{k}e^{iv_gk(t-t_0)}b_k(t_0)$. For a coherent state input $|E_p,\omega_p\ra$ from the left of the 2LE, we have $a_{k}(t_0)|E_p,\omega_p\ra=(\sqrt{\mathcal{L}}E_p/v_g) \delta_{k,\om_p/v_g}|E_p,\om_p\ra$ and $b_{k}(t_0)|E_p,\omega_p\ra=0$, where $\delta_{k,\om_p/v_g}$ is a Kronecker delta function. We take expectation of the operators in Eqs.~\ref{sig1}, \ref{sigcor1} in $|E_p,\omega_p\ra$, and rewrite these equations as
\bea
\f{d\mathcal{S}_1(t)}{dt}&=&(i\Delta_p-2\Gamma_p)\mathcal{S}_1(t)-i\Omega_p+2i\Omega_p \mathcal{S}_2(t),\\
\f{d\mathcal{S}_2(t)}{dt}&=&-4\Gamma_p\mathcal{S}_2(t)+i\Omega_p(\mathcal{S}_1(t)-\mathcal{S}_1^*(t)),
\eea 
where $\Omega_p=\tilde{g}_pE_p/v_g$, $\mathcal{S}_1(t)=\la E_p,\omega_p|\sigma(t)|E_p,\omega_p\ra e^{i\omega_p(t-t_0)}$ and $\mathcal{S}_2(t)=\la E_p,\omega_p|\sigma^{\dg}(t)\sigma(t)|E_p,\omega_p\ra$. The coupled differential equations of $\mathcal{S}_1(t),\mathcal{S}_1^*(t)$ and $\mathcal{S}_2(t)$ can be solved for some initial conditions of these variables, and the long-time steady-state solutions are independent of the initial conditions for the 2LE. We find at steady-state:
\bea
\mathcal{S}_1(t \to \infty)=\f{i\Omega_p(-i\Delta_p-2\Gamma_p)}{\Delta_p^2+4\Gamma_p^2+2\Omega_p^2},~\mathcal{S}_2(t\to \infty)=\f{\Omega_p^2}{\Delta_p^2+4\Gamma_p^2+2\Omega_p^2},\nn
\eea  
which can be used to determine the reflection current $\mathcal{J}_c$ at steady-state for a coherent state input:
\bea
\mathcal{J}_c&=&i\tilde{g}_p \la E_p,\omega_p|(\sigma^{\dg}(t)\tilde{b}_0(t)-\tilde{b}_0^{\dg}(t)\sigma(t))|E_p,\omega_p \rangle\nn\\&=&2\Gamma_p\mathcal{S}_2(t\to \infty)=\f{2\Gamma_p\Omega_p^2}{\Delta_p^2+4\Gamma_p^2+2\Omega_p^2}=\f{\Omega_p^2}{2v_g\Gamma_p}\f{4v_g\Gamma_p^2}{\Delta_p^2+4\Gamma_p^2}-\Big(\f{\Omega_p^2}{2v_g\Gamma_p}\Big)^2\f{16v_g^2\Gamma_p^3}{(\Delta_p^2+4\Gamma_p^2)^2}+\mathcal{O}(\Omega_p^6), \label{refCurCoh}
\eea
where the last expansion in order of $\Omega_p^2$ is obtained for a weak coherent state input, i.e., $\Omega_p^2/(2\Gamma_p^2)\ll1$. We can further identify $I_{cp}=\la E_p,\om_p| \sum_{k} a_{k}^{\dg}(t_0)a_{k}(t_0)|E_p,\om_p\ra / \mathcal{L}=E_p^2/v_g^2=\Omega_p^2/(2v_g\Gamma_p)$.

Taking Fourier transform to real space, we get from Eq.~\ref{af}
\bea
\tilde{a}_x(t)=\f{1}{\sqrt{\mathcal{L}}}\eta_a\big(t-\f{x}{v_g}\big)-\f{i\bar{g}_p}{v_g}\sigma\big(t-\f{x}{v_g}\big)\theta(x)\theta(v_gt-x),
\eea
where we set $t_0=0$. Thus, we find for the first-order coherence:
\bea
G^{(1)}_c(x',x;t)&=&\la E_p,\om_p|\tilde{a}_{x'}^{\dg}(t)\tilde{a}_{x}(t)|E_p,\om_p \ra \nn\\
&=&I_{cp}e^{i\om_p(x-x')/v_g}+\f{\bar{g}_p^2}{v_g^2}\la E_p,\om_p|\sigma^{\dg}\big(t-\f{x'}{v_g}\big)\sigma\big(t-\f{x}{v_g}\big)|E_p,\om_p \ra\theta(x)\theta(x')\theta(v_gt-x')\theta(v_gt-x)\nn\\
&-&\f{i\bar{g}_pE_p}{v_g^2}\la E_p,\om_p|\sigma\big(t-\f{x}{v_g}\big)|E_p,\om_p \ra e^{i\om_p(t-x'/v_g)}\theta(x)\theta(v_gt-x)\nn\\
&+&\f{i\bar{g}_pE_p}{v_g^2}\la E_p,\om_p|\sigma^{\dg}\big(t-\f{x'}{v_g}\big)|E_p,\om_p \ra e^{-i\om_p(t-x/v_g)}\theta(x')\theta(v_gt-x').
\eea
For $x'<0,x>0$, the above expression becomes
\bea
G^{(1)}_c(x'<0,x>0;t)&=&\f{E_p^2}{v_g^2}e^{i\om_p(x-x')/v_g}\Big(1-\f{2i\Gamma_p}{\Omega_p}\mathcal{S}_1\big(t-\f{x}{v_g}\big)\Big).
\eea
At very long time when $t\gg x/v_g$, we can replace $\mathcal{S}_1(t-x/v_g)$ by $\mathcal{S}_1(t \to \infty)$ to obtain 
\bea
G^{(1)}_c(x'<0,x>0;t)&=&I_{cp}e^{i\om_p(x-x')/v_g}\f{i\Delta_p}{i\Delta_p-2\Gamma_p}\Big(1-\f{8v_g\Gamma_p^2I_{cp}}{i\Delta_p(\Delta_p^2+4\Gamma_p^2)}+\mathcal{O}(\Omega_p^4)\Big),
\eea
where we again employ an expansion in $\Omega_p^2$ for a weak coherent state input.
\end{document}